\documentclass[final,5p,times,twocolumn]{elsarticle}

\usepackage{hyperref,amsmath,amssymb}

\usepackage[table]{xcolor}
\usepackage{url}
\definecolor{Gray}{gray}{0.9}

\journal{Nuclear Instruments and Methods A}

\begin{document}

\begin{frontmatter}

\title{Inverse Low Gain Avalanche Detectors (iLGADs) for precise tracking and timing applications}


\cortext[mycorrespondingauthor]{Corresponding author}
\author[a,b]{E. Curr\'as\corref{mycorrespondingauthor}}
\ead{ecurrasr@cern.ch}
\author[c]{M. Carulla}
\author[a]{M. Centis Vignali}
\author[a,b]{J. Duarte-Campderros}
\author[a,b]{M. Fern\'andez}
\author[c]{D. Flores}
\author[b]{A. Garc\'ia}
\author[b]{G. G\'omez}
\author[b]{J. Gonz\'alez}
\author[c]{S. Hidalgo}
\author[b]{R. Jaramillo}
\author[c]{A. Merlos}
\author[a]{M. Moll}
\author[c]{G. Pellegrini}
\author[c]{D. Quirion}
\author[b]{Iv\'an Vila}

\address[a]{CERN, Organisation europ\'enne pour la recherche nucl\'éaire, CH-1211 Genéve 23, Switzerland}
\address[b]{Instituto de F\'isica de Cantabria (CSIC-UC), Avda. los Castros s/n, E-39005 Santander, Spain}
\address[c]{Instituto de Microelectr\'onica de Barcelona (IMB-CNM-CSIC), Bellaterra (Barcelona), 08193, Spain}

\begin{abstract}
Low Gain Avalanche Detector (LGAD) is the baseline sensing technology of the recently proposed Minimum Ionizing Particle (MIP) end-cap timing detectors (MTD) at the Atlas and CMS experiments. The current MTD sensor is designed as a multi-pad matrix detector delivering a poor position resolution, due to the relatively large pad area, around 1 $mm^2$; and a good timing resolution, around 20-30 ps. Besides, in his current technological incarnation, the timing resolution of the MTD LGAD sensors is severely degraded once the MIP particle hits the inter-pad region since the signal amplification is missing for this region. This limitation is named as the LGAD fill-factor problem.
To overcome the fill factor problem and the poor position resolution of the MTD LGAD sensors, a p-in-p LGAD (iLGAD) was introduced. Contrary to the conventional LGAD, the iLGAD has a non-segmented deep p-well (the multiplication layer). Therefore, iLGADs should ideally present a constant gain value over all the sensitive region of the device without gain drops between the signal collecting electrodes; in other words, iLGADs should have a 100${\%}$ fill-factor by design. In this paper, tracking and timing performance of the first iLGAD prototypes is presented.

\end{abstract}

\begin{keyword}
\texttt LGAD, iLGAD, Timing Detector, gain, jitter, microstrip
\end{keyword}

\end{frontmatter}

\section{Introduction}
The high-luminosity upgrade of the Large Hadron Collider (HL-LHC) is foreseen to start in 2026 with a delivery of an integrated luminosity up to 4000\,$fb^{-1}$ during its 10\,years of operation. The HL-LHC will operate at a stable luminosity of $5.0\times10^{34}$\,$cm^{-2}s^{-1}$, with an ultimate scenario of $7.5\times10^{34}$\,$cm^{-2}s^{-1}$. The pileup will be one of the main challenges of the HL-LHC, the interaction region will spread over about 50\,mm in RMS along the beam axis, and produce an average of 1.6\,collisions/mm for an average of 200\,pp interactions per bunch crossing. In these conditions a major challenge is to reject the charged particles produced by the pileup. It is possible to determine if two tracks are coming from a single interaction or from different ones if their time is measured with enough precision.

In this context MIP timing detectors are proposed \cite{Allaire:2302827,Pena:2639962}. Providing a time resolution of 30 ps in the forward region these detectors will be able to mitigate the high pileup and improve the performance of the ATLAS and CMS detectors.

\begin{figure}[t]
\centering
\includegraphics[width=8cm]{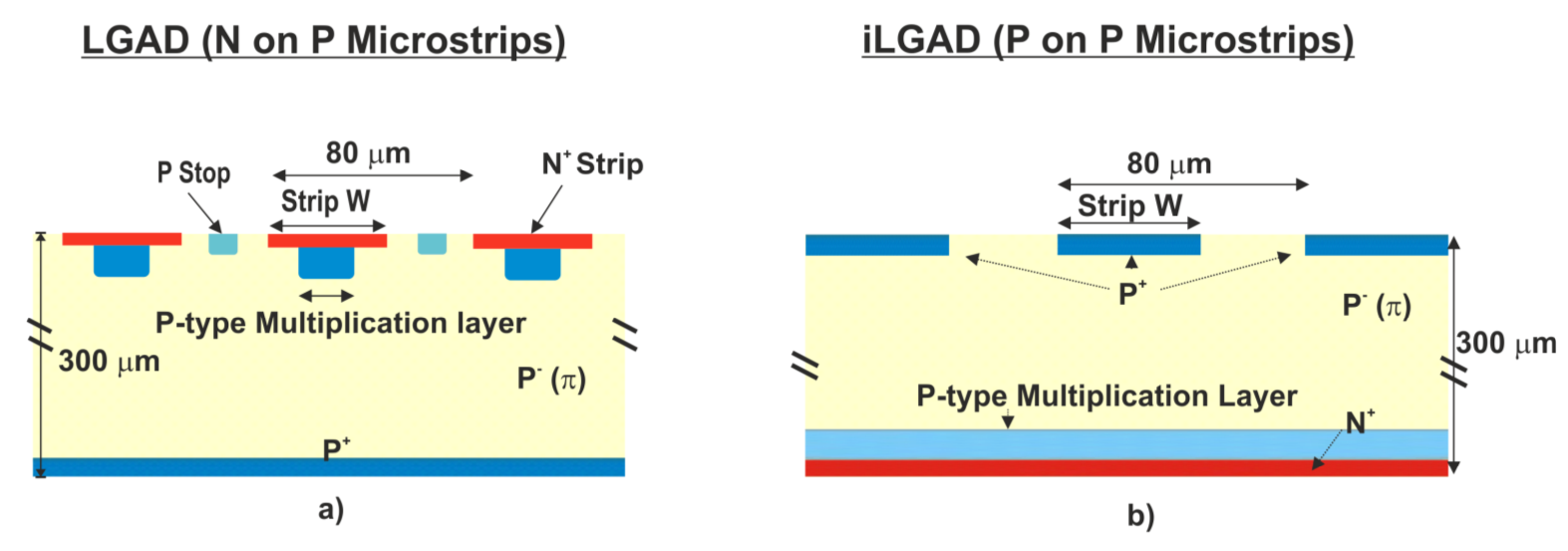}
\caption{Cross-sections of the core layout of LGAD (left) and iLGAD (right) microstrip designs \cite{PELLEGRINI201412}.} 
\label{cross-section}
\centering
\end{figure}

The MTD sensors will be made of Low Gain Avalanche Detectors (LGAD) \cite{PELLEGRINI201412,CARTIGLIA2015141}. LGADs are n-on-p silicon detectors with an internal gain. To obtain this gain an extra, highly doped, p-layer is added just below the p-n junction of a PIN diode. This highly doped region will create a very high electric field region. This electric field will induce an avalanche multiplication of the electrons and thus create additional electron-hole pairs. The LGAD structure is designed to exhibit a moderate gain with an almost linear evolution in a wide range of reverse voltage values. Moreover, microstrip and pixel detector layouts with fine segmentation pitches can be easily obtained with the LGAD approach with a high SNR value, when LGAD is compared with p-i-n detector. Therefore, precise measurements of position and time of arrival of the incident particles can be achieved with LGAD designs.
The timing and position resolution of the LGAD sensors are severely degraded whenever a MIP particle hits the inter-pad region since the signal amplification is missing for this region, as can be seen on figure\,\ref{cross-section}. This limitation is named the LGAD fill-factor problem.

To address this problem on the LGAD sensors, a p-in-p LGAD (iLGAD) was introduced. Contrary to the conventional LGAD design, the iLGAD has a non-segmented multiplication layer, and it should ideally present a constant gain value over all the sensitive region of the device without gain drops between the signal collecting electrodes, see figure\,\ref{cross-section}.
We have experimentally confirmed this feature on a strip-like segmented iLGAD and compare it against a conventional strip-like LGAD and PIN devices. First studies on timing and tracking performance of the first iLGAD prototype are presented here.

\section{Sensor prototype description}
In this work two different types of microstrip detectors fabricated at IMB-CNM (CSIC) are studied \cite{Carulla_2016}. 

Microstrip LGAD detectors were initially designed to be fully compatible with the standard LGAD technology. The core LGAD microstrip schematic cross-section is plotted on figure\,\ref{cross-section} (left). Three strips are included with a P-stop diffusion in between to provide isolation. n-on-p microstrips are implemented with a shallow $n^+$ diffusion overhanging the p-type multiplication diffusion. LGAD microstrips were fabricated on a 285\,$\mu$m thick high resistivity FZ wafer, with a total detection area of 1\,$cm^2$ and two different strip layouts: pitch of 80\,$\mu$m (50\,$\mu$m of strip width), $n^+$ diffusion of 32\,$\mu$m and multiplication diffusion of 20 $\mu$m and pitch of 160\,$\mu$m (130\,$\mu$m of strip width), $n^+$ diffusion of 112\,$\mu$m and multiplication diffusion of 100\,$\mu$m.

The iLGAD structure is also based on the conventional LGAD process technology but with the difference that the segmentation is now located at the $p^+$ side, according to the schematic cross-section of the core iLGAD depicted on figure\,\ref{cross-section} (right). Therefore, the multiplication diffusions are no longer locally performed and the gain is the same through the strip providing a position-sensitive detector with uniform amplification wherever a particle hits the detector.

Standard PIN microstrip detectors were produced for comparison too.

\section{Tracking: the LGAD fill-factor problem}

The tracking performance of one LGAD and one iLGAD strip detector was studied in a test beam at CERN-SPS and compared with a standard PIN strip detector \cite{vila:trento,jordi:RD50}. These three strip detectors were unirradiated and consisted of 45\,strips with a 160\,$\mu$m pitch. The read out was done using the ALIBAVA DAQ. An EUDET-type beam telescope was used for the tracks reconstruction. All measurements were performed at room temperature.

\begin{figure}[t]
\centering
\includegraphics[width=7cm]{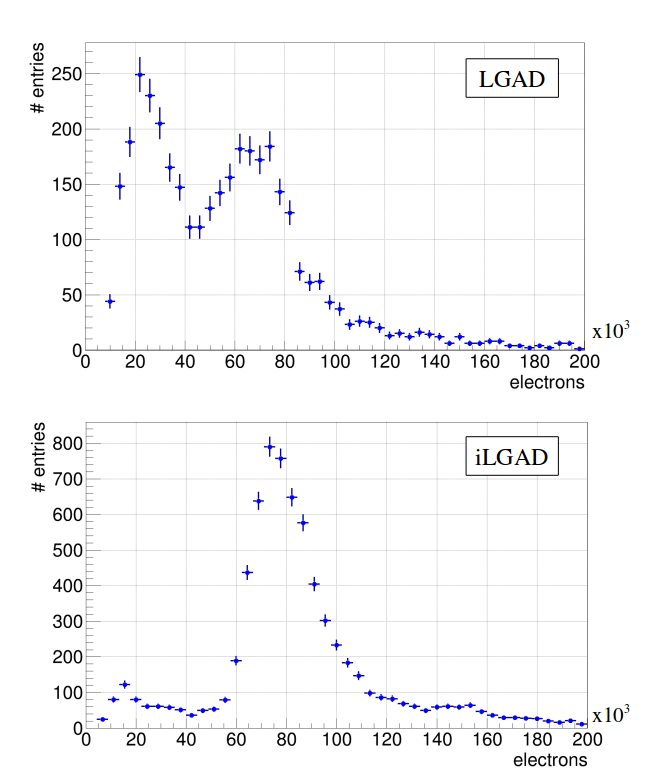}
\caption{Charge distribution measured during the test beam for one LGAD strip detector (top) and one iLGAD strip detector (bottom). The LGAD fill-factor problem is easily spotted and on the contrary, it is not present in the iLGAD structure.} 
\label{fill-factor}
\centering
\end{figure}

The big advantage of the iLGAD technology was confirmed during the test beam. It was proved that while in the LGAD strip detector the signal is severely degraded in the inter-pad region, the iLGAD presents a very constant gain value over all the sensitive region of the device. These results are shown on figure\,\ref{fill-factor}. We see that the charge distribution of the LGAD measured during the test beam presents two peaks. One around 24\,ke, that corresponds to the MIP particles that cross the interstrip region and the signal that they generate is not amplified (same charge measure in the PIN strip). The second peak is around 77\,ke, and it corresponds to the particles that cross the region where the signal is amplified. Notice that the amplification expected for these sensors operated at a voltage of 120\,V is around 3. On the other hand, the same plot produced for the iLGAD detector presents only one peak in the charge distribution around 75\,ke. In this case,  the signals produced for all the MIP particles that cross the sensitive region of the device are amplified, resulting in a much better and uniform response along the sensitive region.

\begin{figure}[t]
\centering
\includegraphics[width=8cm]{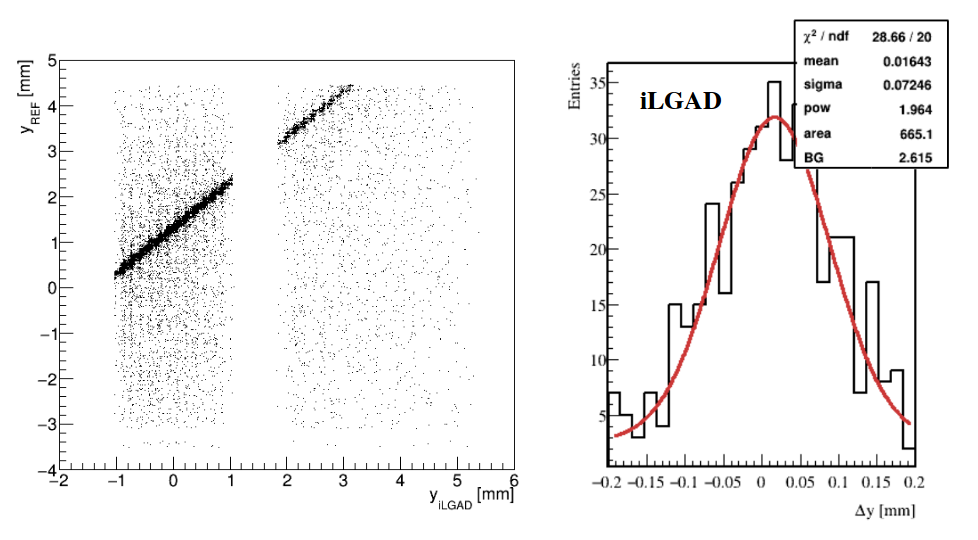}
\caption{Correlations between the reference sensor and the iLGAD sensor during the test beam (left). Spatial resolution measured on the iLGAD sensor at 300\,V and at room temperature (right).}
\label{spatial_resolution}
\centering
\end{figure}

In addition, the spatial resolution was measured for the iLGAD detector. On figure\,\ref{spatial_resolution} (left) the correlations obtained between the strip reference detector and the iLGAD strip detector are shown. The empty columns between 1 and 2\,mm are due to non working strips on the iLGAD detector. On the same figure\,\ref{spatial_resolution} (right) the spatial resolution measured on the iLGAD strip detector at 300\,V is shown. Because of a non optimal read out system, where the electronics front-end was not ready to read large signals presenting saturation above 100\,V, the value presented here of 72\,$\mu$m is not the optimal.

\section{Timing}

After proving that the iLGAD structure presents a 100${\%}$ fill-factor, the timing capabilities of these detectors were studied in the laboratory. The objectives and results of this part are described bellow.

\subsection{Setup description}

\begin{figure}[t]
\centering
\includegraphics[width=7cm]{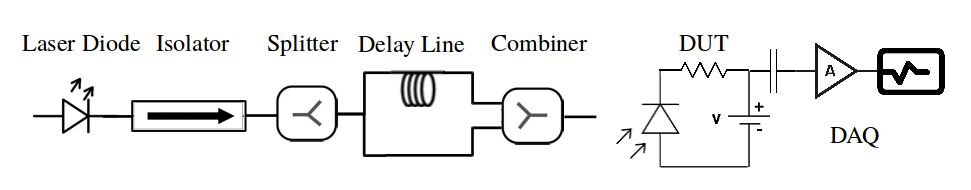}
\caption{Timing setup schematic with all its main components: laser diode, isolator, splitter, delay line, combiner and DUT.} 
\label{timing-setup}
\centering
\end{figure}

A dedicated timing setup was built for this purpose. The main idea was to avoid any external time reference in order to reduce the uncertainty in the measurements coming from its time resolution. The schematic of the setup is shown on figure\,\ref{timing-setup}. We used a picosecond pulsed infrared laser head (1060\,nm). Each laser pulse is split in two lines, in one of which a fixed delay was introduced. Then, these two lines are recombined in one line that illuminates the Device Under Test (DUT). In this way, we have a fixed time interval between laser pulses arriving to the DUT and the use of an external reference is not needed. The time interval between pulses is $\sim$52\,ns. The DUT signal is amplified using a miteq\,1660 current amplifier with a gain of 60\,dB and then this signal is digitized with a 25\,GSa/s oscilloscope.

\begin{figure}[t]
\centering
\includegraphics[width=7cm]{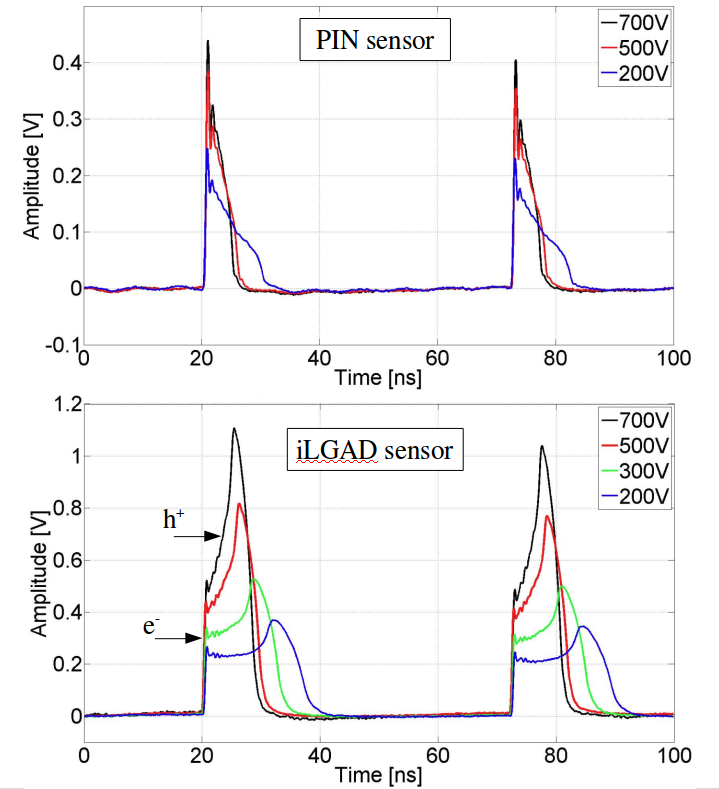}
\caption{Two examples of waveforms from our DUTs at different voltages. On the top one is shown the PIN strip detector and in the bottom one the iLGAD strip detector. In the iLGAD pulses it can be seen the contribution to the signal of the primary electrons and the secondary holes coming from the amplification.} 
\label{waveforms}
\centering
\end{figure}

For the subsequent analysis some parameters are measured from each collected waveform: rise time, signal amplitude, charge and noise. The charge is estimated by the integral of the transient pulse. The gain is calculated as the ratio between the iLGAD collected charge over the PIN collected charge. The time resolution ($\sigma_{t}$) is obtained from the dispersion (standard deviation) of the time difference between the two pulses divided by $\sqrt{2}$, since both pulses should contribute equally to the time resolution. Moreover, the time resolution is going to be dominated by the jitter component, since the pulse shape and amplitude are very stable. Two examples of the pulse's waveforms at different bias voltages are shown on figure\,\ref{waveforms}, on the top graph for the PIN strip detector and on the bottom graph for the iLGAD strip detector. To show them with more clarity, the signals were averaged. For this part we used: a p-in-p PIN strip detector with a pitch of 80\,$\mu$m and 50\,$\mu$m p+ implant width; and an iLGAD strip detector with a pitch of 160\,$\mu$m and an ohmic contact width of 130\,$\mu$m. The laser spot is confined inside the strip width, inside an optical window where no Al was deposited on top of the ohmic contact.

\subsection{Constant Fraction Discrimination method (CFD)}
\begin{figure}[t]
\centering
\includegraphics[width=8cm]{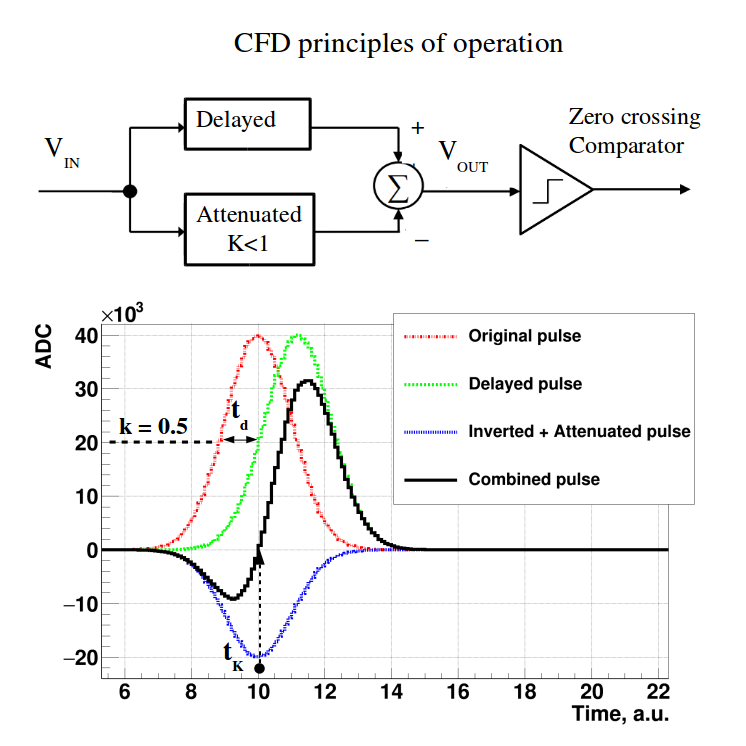}
\caption{Scheme of CFD principle of operation. In this example an attenuation value of k = 0.5 has been chosen. $t_{d}$ represents the time that the original signal is delayed and $t_{k}$ would be the reference time extracted with this method.}
\label{cfd-method}
\centering
\end{figure}

One of our goals is to emulate by software a real CFD electronic circuit, and the analysis of the data has been done in this way. CFD is a technique developed to provide information about the arrival time of an event with no dependency on the amplitude of the signal. The principle of operation is based on detecting the zero crossing of a bipolar signal obtained by subtracting a fraction of the input signal ($0 < k < 1$) to its delayed copy as it is illustrated on figure\,\ref{cfd-method}. This bipolar signal crosses the baseline at a fixed time ($t_{k}$) with respect to the start of the original signal. 

\subsection{Results}

\begin{figure}[t]
\centering
\includegraphics[width=8cm]{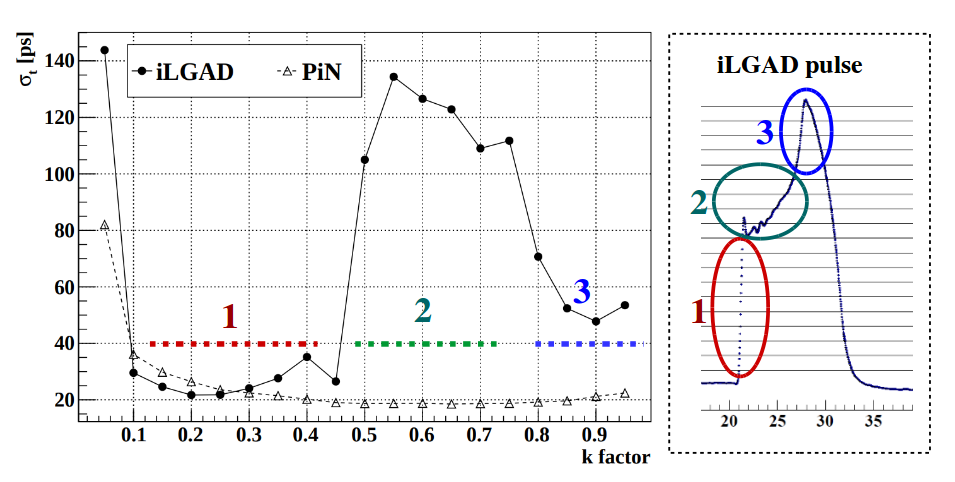}
\caption{In the left hand plot is shown the time resolution measured as a function of the attenuation factor (k) at 700\,V. For the PIN detector any value of k gives the optimal time resolution; but in the case of the iLGAD detector there are three different zones related to the three different zones labeled in the pulse in the right hand plot.}
\label{sigma_k}
\centering
\end{figure}

As it was introduced before, one parameter that can be tuned in this CFD method is the attenuation factor of the original signal (k). The optimal k value will depend on the shape of the signal to be analyzed. For a given pulse shape the delay of the original signal is determined by the k value; the signal delay $(t_{d})$ is the difference between the time of its maximum and the time where the signal crosses the fractional threshold (k). Thus,  in order to check what will be the optimal k value in our system, different k values were tested for both detectors. The time resolution obtained $(\sigma_t)$ is plotted as a function of this attenuation factor (k), the results are shown on figure\,\ref{sigma_k}. 

It can be seen that in the case of the PIN strip detector any value of k will give the optimal value of the time resolution. Meanwhile, in the case of the iLGAD strip detector, three different zones can be observed, they can be identified with the shape of the pulse already described. The zone labeled as 1 is mainly influenced by the fast electron collection and it gives the optimal value in terms of time resolution. In this zone, the time resolution measure on the iLGAD detector reach a value of 20\,ps, very similar to the one measured on the PIN detector. In the zone labeled as 2, the secondary holes start to contribute to the signal and we have a much more ripply and slower edge section, which implies a worse time resolution. Finally, close to the maximum of the signal, zone 3, we have a stepper pulse edge with a faster slewing rate and in consequence a better time resolution. These measurements where done at 700\,V and a room temperature, where the iLGAD presented a gain value of 4.8. The amount of primary carries generated by the laser pulse were similar for both devices.

\section{Conclusions and Outlook}
\begin{figure}[t]
\centering
\includegraphics[width=8.5cm]{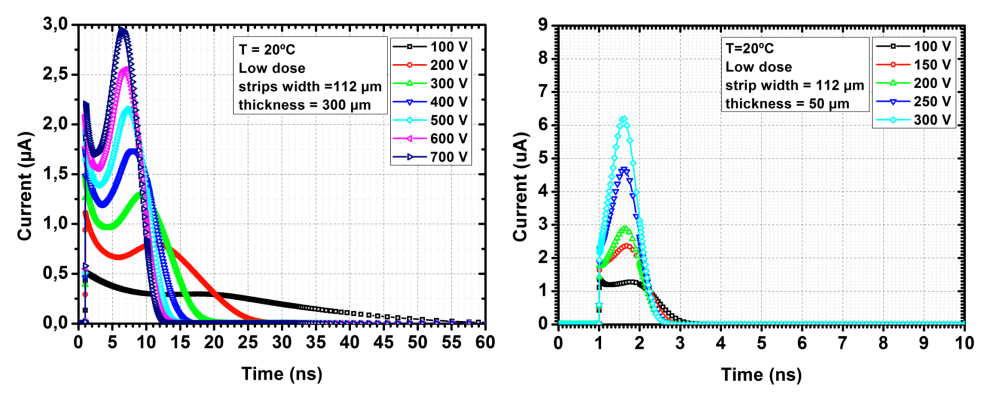}
\caption{Transient TCAD simulation from the iLGAD sensor studied in this work (left) compared to the responds of a similar 50\,$\mu$m thick strip iLGAD (right). Notice the different time range of the horizontal axis.}
\label{simulation}
\centering
\end{figure}

In this work a detailed characterization of iLGAD strip detectors were presented. It is a very promising candidate to become a true 4-dimensional tracking technology. Test beam measurements demonstrated the homogeneity in the amplification over all the sensitive region of the device without gain drops between the signal collecting electrodes.
Moreover, a comparison between the timing performance of one iLGAD strip detector with a similar PIN strip detector was presented. The time resolution was estimated using a dedicated laser setup without an external reference and it was computed emulating an electronic CFD method. Simulations show that these promising results can be improved further if thinner sensors are used. As shown in figure\,\ref{simulation}, they benefit from a much smaller rise time of the signal maintaining a good SNR. This feature will improve the time resolution with respect to the thicker devices studied here.

\section*{Acknowledgments}

RD50 Collaboration for its support. This activity was partially supported by the Spanish Ministry of Science under grants FPA2015-71292-C2-2-P, FPA2017-85155-C4-1-R and FPA2017-85155-C4-2-R; and the European Union Horizon 2020 Research and Innovation Programme under Grant Agreement no. 654168 (AIDA-2020).
 
\bibliography{bibliography}

\end{document}